\documentclass[preprint,showpacs,preprintnumbers,amsmath,amssymb]{revtex4}

\usepackage{graphicx}
\usepackage{dcolumn}
\usepackage{bm}

\newcommand{\eref}[1]{(\ref{#1})}

\begin{document}
\title{Surface-Invariants in 2D Classical Yang-Mills Theory}
\author{Rafael D\'{\i}az}
 \email{rdiaz@euler.ciens.ucv.ve}
 \affiliation{Escuela de Matem\'aticas, Facultad de Ciencias, Universidad Central de Venezuela, AP 20513,
Caracas 1020-A, Venezuela}
\author{E. Fuenmayor}%
 \email{efuenma@fisica.ciens.ucv.ve}
\author{Lorenzo Leal}
 \email{lleal@fisica.ciens.ucv.ve}
\affiliation{Centro de F\'{\i}sica Te\'orica y Computacional,
Facultad de Ciencias, Universidad Central de Venezuela, AP 47270,
Caracas 1041-A, Venezuela}%

\date{\today}

%

\begin{abstract}

We study a method to obtain invariants under area-preserving
diffeomorphisms associated to closed curves in the plane from
classical Yang-Mills theory in two dimensions. Taking as starting
point the Yang-Mills field coupled to non dynamical particles
carrying chromo-electric charge, and by means of a perturbative
scheme, we obtain the first two contributions to the on shell
action, which are area-invariants. A geometrical interpretation of
these invariants is given.

\end{abstract}

\maketitle

\section{Introduction}

Quantum topological field theories, such as the Chern-Simons  and
$BF$ models, are gauge-invariant and metric-independent
 theories. Due to the latter
characteristic, they can be used for the study of knot and link
invariants in  three dimensional ($3d$) spaces
\cite{witten,labastida,guadagnini}, and for the study of
generalizations of these invariants appropriated to the dimension
of the base manifold.

This relationship between topological theories and knot invariants
subsists even at the classical level, as has been shown in the
Abelian case \cite{L3}, and also in the non-Abelian case
\cite{Nudos} by means of a perturbative study of the classical
equations of motion of topological theories coupled to external
point-particles carrying non-Abelian charge (Wong particles
\cite{wong}). The method presented in these references rests upon
the fact that the classical action of the theory must retain its
diffeomorphism-invariant character when it is evaluated on-shell.
Hence, the on-shell action of the Chern-Simons (or $BF$) theories
coupled in a suitable manner to particles should  yield link
invariants in $3d$, just as the vacuum expectation value of the
Wilson-Loop does within the quantum field approach. This result
concerning the classical perturbative treatment can be rigorously
proven and generalized to situations where the symmetry group is
other than the group of diffeomorphisms  of the base manifold
\cite{rafael}.

The purpose of this article, within the context of the ideas and
procedures developed in references \cite{L3,Nudos,rafael}, is to
apply the classical perturbative method outlined above to $2d$
Euclidean Yang-Mills theory, which yields and example of how the
method works when the symmetry group is smaller than the full
group of diffeomorphism. As it is well known, $2d$ Yang-Mills is
invariant under diffeomorphisms that preserve areas
\cite{y1,y2,y3}, which reflects in the fact that the Wilson loop
exhibits  an exponential dependence on the areas of the loop
\cite{y4}.  $2d$ Yang-Mills theories have been studied in many
contexts. For instance, they describe closed topological strings
with bounded states, and q-deformed $2d$ Yang Mills theories in
Riemann surfaces give topological invariants in one dimension
higher \cite{y5,y6}. Furthermore, under certain conditions,
Euclidean $2d$ Yang-Mills theory  mimics the confining phase of
$QCD$ to a good degree of accuracy \cite{y7,y8}.

The paper is organized as follows. In Section II we  apply the
general method developed in \cite{L3,Nudos,rafael} to $2d$
Yang-Mills theory coupled to external Wong-particles. After
presenting the model we discuss a perturbative scheme  for solving
the classical equations of motion, and calculate the on-shell
action to the first two orders in the coupling constant. In
section \ref{sec3} we find a geometric interpretation of the
invariants. We also discuss the consistency conditions that gauge
invariance imposes over the classical equations of motion, and
their relationship with the definition of the surface-invariants.
To this end, we perform a gauge transformation that amounts to
attaching a bundle of straight lines to each point of space. In
this gauge, area-invariants become neatly expressed in terms of
the areas formed by the cross product among the tangent vectors to
these ``fibers'' and those tangent to the curves. Some final
comments close this section.

\section{Yang-Mills theory in 2 dimensions coupled to Wong particles}\label{sec2}

\subsection{The model and the general method}

Our starting point will be the action
\begin{equation}\label{2.1a}
S=S_{YM}+ S_{int}=\frac{1}{2g^{2}}\int d^{2}x
Tr(F_{\mu\nu}F^{\mu\nu})+\sum_{i=1}^{n}\int_{\gamma_{i}}d\tau
Tr(K_{i}g_{i}^{-1}(\tau)D_{\tau}g_{i}(\tau)),
\end{equation}
where the last term corresponds to the interaction of $n$
non-dynamical Wong particles (i.e., classical particles with
cromo-electrical charge) \cite{wong}. $g$ is the gauge coupling
constant of the Yang-Mills field. Since we are dealing with the
Euclidean theory, there is no distinction between ``Greek'' ($\mu,
\nu,$ etc.) sub or super-scripts, so we shall use both
indistinctly. We will use the following conventions for the
$N^{2}-1$ generators $T^{a}$ of the $su(N)$ algebra and for the
gauge field $A_{\mu}$:
\begin{eqnarray}\label{2.1b}
&&Tr(T^{a}T^{b})= -\frac{1}{2} \delta^{ab}\\
&&[T^{a},T^{b}]= f^{abc}T^{c}\\
&&A_{\mu}= A_{\mu}^{a}T^{a}\\
&&A_{i}=A_{\mu}(z_{i}(\tau))\dot{z}^{\mu}_{i}(\tau).
\end{eqnarray}

As dynamical variables we take the potentials  $A^{a}_\mu$ and the
$SU(N)$ matrices in the fundamental representation $g_{i}(\tau)$,
associated with the internal degrees of freedom of the Wong
particles \cite{wong}. Instead, the trajectories
$z^{\mu}_{i}(\tau)$ of the particles are given. The curves
$\gamma_{i}$ drawn by the particles are then taken as closed
curves in the Euclidean plane, surrounding surfaces $\Sigma_{i}$
whose ``area-invariants" we are going to study.

In equation (\ref{2.1a}) we use the covariant derivative along the
world-line of the $i$-particle:
$D_{\tau}g_{i}(\tau)=\dot{g}_{i}+A_{i}(\tau)g_{i}(\tau)$. Also, it
appears $K_{i}\equiv K_{i}^{a}T^{a}$, which is a constant element
of the algebra related to the initial value of the cromo-electric
charge $I_{i}(\tau)$ through
\begin{equation}\label{2.1d}
I_{i}(\tau)\equiv g_{i}(\tau)K_{i}g_{i}^{-1}(\tau).
\end{equation}
Finally, the Yang-Mills field tensor is defined as usual by
\begin{equation}\label{2.2a}
F_{\mu\nu}=\partial_{\mu}A_{\nu}-\partial_{\nu}A_{\mu}+[A_{\mu},A_{\nu}].
\end{equation}

The first term in the action (i.e., the Yang-Mills one) is
invariant under area-preserving diffeomorphisms. The second one is
invariant against arbitrary diffeomorphisms \cite{Nudos,wong}.
Hence, the whole action is invariant under area-preserving
diffeomorphisms. Also, it is invariant under gauge transformations
\begin{eqnarray}\label{2.1c}
A_{\mu}\rightarrow
A^{\Omega}_{\mu}&=&\Omega^{-1}A_{\mu}\Omega+\Omega^{-1}\partial_{\mu}\Omega ,\\
K_{i}\rightarrow K_{i}^{\Omega}&=&K_{i},\\
g_{i}\;\,\rightarrow\;\, g_{i}^{\Omega}&=&\Omega^{-1}g_{i},\\
I_{\i}\;\;\rightarrow\;\; I_{i}^{\Omega}&=&\Omega^{-1}I_{i}\Omega.
\end{eqnarray}

In can be seen that the interaction term of the action
(\ref{2.1a}) may be put in the form
\begin{equation}\label{2.3a}
S_{int}=\sum_{i=1}^{n}\int_{\gamma_{i}}d\tau
\dot{z}_{i}(\tau)\left(Tr[K_{i}g^{-1}_{i}\partial_{\mu}g_{i}]+\int
d^{2}x \delta^{(2)}(x-z_{i}(\tau))
Tr[I_{i}(\tau)A_{\mu}\left(z_{i}(\tau)\right)]\right),
\end{equation}
hence,  varying the action (\ref{2.1a}) with respect to the
dynamical variable $A_{\mu}^{a}$ we obtain
\begin{equation}\label{2.6a}
D_{\mu}F^{\mu\nu}=\Lambda J^{\nu}=\Lambda
\sum_{i=1}^{n}\int_{\gamma_{i}}d\tau
\dot{z}^{\nu}_{i}(\tau)I_{i}(\tau)\delta^{(2)}(x-z_{i}(\tau)),
\end{equation}
where we have set $\Lambda\equiv g^2$.

To write down the equations of motion for the internal variables
$g_{i}(\tau)$ we follow the procedure given in
\cite{balachandran}. Take a parametrization of the group elements
\begin{equation}\label{2.4a}
g_{i}=g_{i}(\xi_{i})=e^{\xi^{a}_{i}T^{a}},
\end{equation}
and perform variations of $S_{int}$ with respect to the $N^{2}-1$
independent parameters $\xi^{a}_{i}(\tau)$. This leads us to the
Euler-Lagrange equations
\begin{equation}
\frac{\partial L}{\partial \xi^{a}_{i}(\tau)} -
\frac{d}{d\tau}\Big(\frac{\partial L}{\partial
\dot\xi^{a}_{i}(\tau)}\Big) =0, \label{euler lagrange}
\end{equation}
where we have defined
\begin{equation}
L \equiv \sum_{i} Tr (K_i g^{-1}_{i}(\tau) D_{\tau}g_{i}(\tau)).
\label{lagrangiano}
\end{equation}

It can be seen that equations \eref{euler lagrange} are equivalent
to the gauge-covariant conservation of the non-Abelian charge of
each particle along its world line\cite{balachandran} (this
conservation-law arises by taking the covariant derivative on both
sides of equation (\ref{2.6a}))
\begin{equation}
D_{\tau}I_{i}= \dot{I}_{i} + [A_{i},I_{i}]=0, \label{conservacion}
\end{equation}
whose solution is
\begin{equation}
I_{i}({\tau})= U_{i}({\tau}) \, I_{i}(0)\,U^{-1}_{i}({\tau}) .
\label{solucion I}
\end{equation}
Here $U_{i}({\tau})$ is the time ordered exponential of the gauge
field along  the curve $\gamma_{i}$
\begin{equation}
U_{i}({\tau})= \mathbf{T}\exp \,\,( -\int_{0}^{\tau}
A_{i}({\tau}')\, d\tau' \,\,) . \label{tiempo ord}
\end{equation}
From (\ref{2.1d}) and (\ref{solucion I}) we obtain
$g_{i}(\tau)=U_{i}(\tau)g_{i}(0)$, which in turn implies
$D_{\tau}g_{i}(\tau)=0$. Therefore, we find that the interaction
term $S_{int}$ of the action vanishes  when it is evaluated
on-shell. Hence, we only have to consider the Yang-Mills action
evaluated on the equations of motion.

Plugging (\ref{solucion I}) and (\ref{tiempo ord}) into
(\ref{2.6a}) yields an equation for the gauge potentials in terms
of the curves $\gamma_{i}$ \cite{Nudos}.  Inserting the solution
to this equation into the action (\ref{2.1a}) one would finally
obtain a functional $S([\gamma];\Lambda)$ that only depends on the
curves $\gamma$ and the coupling constant $\Lambda$ \cite{Nudos}.
Following the general arguments discussed in the introduction,
this on-shell action should retain the invariance under
area-preserving diffeomorphisms. Henceforth, the final expression
for $S([\gamma];\Lambda)$, calculated as explained above,
constitutes an invariant under area-preserving diffeomorphisms
associated to the curves $\gamma_{i}$.

\subsection{Perturbative expansion to the first two orders}

Since the equations to be solved are non-linear, we will use a
perturbative method to deal with them, such as in reference
\cite{Nudos}. At last, it will result that the action on-shell may
be written as a power series in the parameter $\Lambda$
\begin{equation}
S_{on-shell}\,( [\gamma_i],\Lambda)= \frac{\Lambda}{2}
\sum_{p=0}^{\infty}\Lambda^p\,\, S^{(p)}[\gamma_i].
\label{Sonshell}
\end{equation}

In order to carry out the perturbative method, we define
quantities:
\begin{eqnarray}\label{2.7a}
a_{\mu}&\equiv &\Lambda^{-1}A_{\mu},\nonumber\\
R^{ac}_{i}&\equiv &
f^{abc}a_{\mu}^{b}(z_{i})\,\dot{z}_{i}^{\mu}\quad \longrightarrow
\quad R^{ac}_{\mu}\equiv f^{abc}a_{\mu}^{b}(z_{i}).
\end{eqnarray}
The equations (\ref{conservacion}) for the parameters, and their
formal solution (\ref{tiempo ord}), adopt the form
\begin{equation}\label{2.8a}
\frac{dI_{i}^{a}(\tau)}{d\tau}\,+\,\Lambda
R^{ac}_{i}(\tau)I_{i}^{c}(\tau)=0
\end{equation}
and
\begin{equation}\label{2.9a}
I_{i}^{a}(\tau)=\big[\mathbf{T}\exp\big(-\Lambda\int_{0}^{\tau}R_{i}(\tau')d\tau'\big)\big]^{ac}
I_{i}^{c}(0),
\end{equation}
respectively.  We need to develop the ordered exponential in the
right hand side of (\ref{2.6a}):
\begin{eqnarray}\label{2.10a}
\Lambda J^{\mu}=&&\,+\Lambda\sum_{i=1}^{n}
\oint_{\gamma_{i}}dz^{\mu}\delta^{(2)}(x-z)\,I_{i}^{a}(0)\nonumber\\
&&-\Lambda^{2}\sum_{i=1}^{n}\oint_{\gamma_{i}}dz^{\mu}\int_{0}^{z}dz_{1}^{\mu_{1}}
R^{aa_{1}}_{\mu_{1}}(z_{1})\,\delta^{(2)}(x-z)\,I_{i}^{a_{1}}(0)\nonumber\\
&&+\Lambda^{3}\sum_{i=1}^{n}\oint_{\gamma_{i}}dz^{\mu}\int_{0}^{z}dz_{1}^{\mu_{1}}
\int_{0}^{z_{1}}dz_{2}^{\mu_{2}}R^{aa_{1}}_{\mu_{1}}(z_{1})R_{\mu_{2}}^{a_{1}a_{2}}(z_{2})
\delta^{(2)}(x-z)I_{i}^{a_{2}}(0)\nonumber\\
&&\vdots\nonumber\\
&&+(-\Lambda)^{p+1}\sum_{i=1}^{n}\oint_{\gamma_{i}}dz^{\mu}\int_{0}^{z}dz_{1}^{\mu_{1}}\ldots\int_{0}^{z_{p-1}}
dz_{p}^{\mu_{p}}R^{aa_{1}}_{\mu_{1}}(z_{1})\nonumber\\
&&\qquad\qquad\qquad\qquad\qquad\qquad\qquad\qquad\ldots
R_{\mu_{p}}^{a_{p-1}a_{p}}(z_{p})
\delta^{(2)}(x-z)I_{i}^{a_{p}}(0)\nonumber\\
&&\vdots
\end{eqnarray}
Also, we expand the ``new'' fields $a_{\mu}$ in powers of
$\Lambda$,
\begin{equation}\label{2.11a}
a_{\mu}^{a}=\sum_{p=0}^{\infty}\Lambda^{p}\,a^{(p)a}_{\mu},
\end{equation}
and introduce them into (\ref{2.10a}) to obtain  the right hand of
(\ref{2.6a}) order by order. Additionally, we have to work out the
left hand side of (\ref{2.6a})
\begin{eqnarray}\label{2.12a}
D_{\mu}F^{\mu\nu}&=&\partial_{\mu}F^{\mu\nu}+[A_{\mu},
F^{\mu\nu}]\nonumber\\
&=&\Box
A^{\nu}-\partial^{\nu}(\partial_{\mu}A^{\mu})+\partial_{\mu}[A^{\mu},A^{\nu}]
+[A_{\mu},\partial^{\mu}A^{\nu}-\partial^{\nu}A^{\mu}]+
\big[A_{\mu}, [A^{\mu},A^{\nu}]\big],
\end{eqnarray}
where $\Box \equiv\partial_{\mu}\partial^{\mu}$ (since we are in
Euclidean space this is the same as the Laplacian $\nabla^{2}$).
Choosing  the Lorentz gauge $\partial_{\mu}A^{\mu}=0$, and
changing to the $a_{\mu}$ variables in (\ref{2.12a}) we can write
the equation of motion (\ref{2.6a}) as
\begin{eqnarray}\label{2.13a}
\Lambda\Box
a^{\nu}_{a}+\Lambda^{2}2a^{\mu}_{b}\partial_{\mu}a^{\nu}_{c}f_{abc}
-\Lambda^{2}a_{\mu\,b}\partial^{\nu}a^{\mu}_{c}f_{abc}+\Lambda^{3}a_{\mu\,
b}a^{\mu}_{d}a^{\nu}_{e}f_{cde}f_{abc}=\Lambda J^{\mu}_{a},
\end{eqnarray}
where the right hand side of this equation is given by
(\ref{2.10a}), with the substitution \eref{2.11a}, as discussed.

We are ready to evaluate the Yang-Mills action on-shell up to the
first orders in $\Lambda$. It will be useful to define the
"Abelian part" of the field tensor
$f_{\mu\nu}\equiv\partial_{\mu}a_{\nu}-\partial_{\nu}a_{\mu}$,
which allows to write the Yang-Mills action as
\begin{eqnarray}\label{2.14a}
S_{YM}&=&\frac{1}{2\Lambda}\int d^{2}x
Tr(F_{\mu\nu}F^{\mu\nu})=\frac{\Lambda}{2}\int d^{2}x
Tr(f_{\mu\nu}+\Lambda[a_{\mu},a_{\nu}])^{2}\nonumber\\&=&\frac{\Lambda}{2}\int
d^{2}x Tr(f_{\mu\nu}f^{\mu\nu}+2\Lambda
f_{\mu\nu}[a_{\mu},a_{\nu}]+\Lambda^{2}[a_{\mu},a_{\nu}][a^{\mu},a^{\nu}]).
\end{eqnarray}

From expression (\ref{2.14a}) it is immediate to obtain the $0-th$
order contribution of the on-shell action (\ref{2.1a})
\begin{eqnarray}\label{2.14b}
S_{YM}^{(0)}&=&\frac{1}{2\Lambda}\int d^{2}x
Tr(F^{(0)}_{\mu\nu}F^{\mu\nu\,(0)})=\frac{\Lambda}{2}\int d^{2}x
Tr(f_{\mu\nu}f^{\mu\nu})=\frac{\Lambda}{2}\int d^{2}x
Tr(\partial_{\mu}a^{(0)}_{\nu}-\partial_{\nu}a^{(0)}_{\mu})^{2},\nonumber\\
\end{eqnarray}
where $a_{\mu}^{(0)}$ is the solution to the $0-th$-order equation
of motion that results from (\ref{2.12a})
\begin{eqnarray}\label{2.15a}
\Box a^{\mu\,(0)}=J^{\mu\,(0)}=\sum_{i=1}^{n}
\oint_{\gamma_{i}}dz^{\mu}\delta^{(3)}(x-z)\,I_{i}^{a}(0).
\end{eqnarray}
Before showing the final expression for the $0-th$ order
invariant, we find it convenient to work out the first order
invariant to the same level of detail. From (\ref{2.14a}) we have
\begin{eqnarray}\label{2.17a}
S_{YM}^{(1)}&=&\int d^{2}x
Tr\left(f^{(1)}_{\mu\nu}f^{\mu\nu\,(0)}+f^{(0)}_{\mu\nu}[a^{\mu\,(0)},
a^{\nu\,(0)}]\right)\nonumber\\&=&\int d^{2}x
Tr\left(-2a_{\nu}^{(0)}\Box
a^{(1)}_{\nu}+f^{(0)}_{\mu\nu}[a^{\mu\,(0)}, a^{\nu\,(0)}]\right),
\end{eqnarray}
where we have thrown away boundary terms and used the Lorentz
gauge. From (\ref{2.12a}), we have that the first order potential
obeys the equation
\begin{eqnarray}\label{2.16a}
\Box
a^{\nu\,(1)}_{a}=a^{\mu\,(0)}_{b}\partial^{\nu}a^{\mu\,(0)}_{c}f_{abc}
-2a^{\mu\,(0)}_{b}\partial_{\mu}a^{\nu\,(0)}_{c}f_{abc}+J^{\nu\,(1)}_{a},
\end{eqnarray}
where $J^{\nu\,(1)}$ can be easily obtained from (\ref{2.10a}).

At this point it is worth noticing that the general structure of
the $p-th$-order  equation of motion is, indeed, the same as that
of the first two ones: the Laplacian of the $p-th$-order potential
$a^{\mu (p)}$ is given in terms of previous orders potentials,
which were already solved as functions of the curves $\gamma_{i}$.
Therefore, the perturbative method can be recursively applied, and
the on-shell action  can be obtained order by order.

Turning back to the first contributions, we see that it remains to
solve the $0-th$ and first orders equations (\ref{2.15a}) and
(\ref{2.16a}), and to substitute these results into (\ref{2.14b})
and (\ref{2.17a}). To this end, we find it convenient to introduce
loop-coordinates \cite{extended}
\begin{eqnarray}\label{2.5}
&&T_{\gamma_{i}}^{\mu_{1}\mu_{2}...\mu_{n}}(x_{1},
x_{2},...,x_{n})=T_{\gamma_{i}}^{\mu_{1}
x_{1}\,\mu_{2}x_{2}...\mu_{n}x_{n}}\equiv
\oint_{\gamma_{i}}dz^{\mu_{1}}\int_{0}^{z}dz_{1}^{\mu_{2}}\times\nonumber\\
&\times &\int_{0}^{z_{1}}dz_{2}^{\mu_{3}} \ldots\int_{0}^{z_{n-1}}
dz^{\mu_{n}}_{n-1}\delta^{(2)}(x_{1}-z)
\delta^{(2)}(x_{2}-z_{1})\delta^{(2)}(x_{3}-z_{2})\ldots\delta^{(2)}(x_{n}-z_{n-1}),\nonumber\\
\end{eqnarray}
that obey the differential constraints
\begin{equation}\label{2.6}
\frac{\partial}{\partial
x_{i}^{\mu_{i}}}T^{\mu_{1}x_{1}\cdots\mu_{i}x_{i}\cdots\mu_{n}x_{n}}=
\left(-\delta(x_{i}-x_{i-1})+\delta(x_{i}-x_{i+1})\right)
T^{\mu_{1}x_{1}\cdots\mu_{i-1}x_{i-1}\,\mu_{i+1}x_{i+1}\cdots\mu_{n}x_{n}},
\end{equation}
where both $x_{0}$ and $x_{n+1}$ are taken as the origin of the
loop. Loop-coordinates also obey the algebraic constraints
\begin{equation}\label{2.7}
T^{\{\mu_{1}\cdots\mu_{k}\}\mu_{k+1}\cdots\mu_{n}}=\sum_{P_{k}}T^{P_{k}(\mu_{1}\cdots\mu_{n})}
=T^{\mu_{1}\cdots\mu_{k}}T^{\mu_{k+1}\cdots\mu_{n}}.
\end{equation}
The sum in the right hand side of the last equation is made over
all the permutations of the indices $\mu$ that preserve the order
on the subset $\mu_{1}\cdots\mu_{k}$ and on the remaining subset
$\mu_{k+1}\cdots\mu_{n}$.

We shall use a generalized Einstein convention
\begin{eqnarray}\label{2.2}
A_{\mu x}B^{\mu x\,\nu y...}&\equiv &\sum_{\mu}\int A_{\mu
x}B^{\mu x\,\nu y...}d^{2}x=\sum_{\mu}\int A_{\mu}(x)B^{\mu\nu
...}(x,y...)d^{2}x.
\end{eqnarray}
A bar over a``continuous index'' breaks the Einstein convention
\begin{equation}\label{2.3}
A_{\mu x\,\nu\bar{y}}B^{\mu x\,\nu\bar{y}}\equiv \sum_{\mu}\int
A_{\mu x\,\nu\bar{y}}B^{\mu x\,\nu\bar{y}}d^{2}x.
\end{equation}

For the sake of simplicity, we restrict ourselves to consider
$SU(2)$ as gauge group. We shall use arrows to denote iso-vectors,
and employ  the "dot" and "cross" product notation.

Using these tools, $S^{(0)}$ and $S^{(1)}$ can be written down as
\begin{eqnarray}\label{2.7d}
S^{(0)}_{YM}&=&2\int d^{2}x\partial_{\mu}\vec{a}^{(0)}_{\nu}
\cdot\vec{f}^{(0)\mu\nu}\nonumber\\&=&-2\,\vec{a}^{(0)}_{\nu
x}\cdot\Box \vec{a}^{(0)}_{\nu x},
\end{eqnarray}
and
\begin{equation}\label{2.8}
S^{(1)}_{YM}=\vec{a}^{(0)}_{\nu x}\cdot \vec{J}^{(1)\nu x}+\frac12
\left(\vec{a}^{(0)}_{\mu x}\times\vec{a}^{(0)}_{\nu x}\right)\cdot
\partial^{\mu}\vec{a}^{(0)\nu x},
\end{equation}
while the $0-th$ and $1-st$ order currents are given by
\begin{eqnarray}\label{2.9}
\vec{J}^{(0)\mu x}&=&\sum_{i}T_{i}^{\mu x}\vec{I}_{i}\nonumber\\
\vec{J}^{(1)\mu x}&=&\sum_{i}T_{i}^{\mu x\,\nu
y}\vec{a}^{(0)}_{\nu y}\times\vec{I}_{i}.
\end{eqnarray}
In turn, equations (\ref{2.15a}) and (\ref{2.16a}) for the zero
and first orders can be written as
\begin{equation}\label{2.10}
\Box \vec{a}^{(0)\mu x}=\sum_{i}T_{i}^{\mu x}\vec{I}_{i},
\end{equation}
and
\begin{equation}\label{2.10.a}
\Box \vec{a}^{\nu
x\,(1)}=\vec{a}^{\mu\bar{x}\,(0)}\times\partial^{\nu}\vec{a}^{\mu\bar{x}\,(0)}
-2\vec{a}^{\mu\bar{x}\,(0)}\times\partial_{\mu}\vec{a}^{\nu\bar{x}\,(0)}+\sum_{i}T_{i}^{\nu
x\,\mu y}\vec{a}^{(0)}_{\mu y}\times\vec{I}_{i}.
\end{equation}
The solution of (\ref{2.10}) is given by
\begin{equation}\label{2.11}
\vec{a}^{(0)\mu x}=G_{x,y}\vec{J}^{(0)\mu
y}=\frac{1}{2\pi}\sum_{i}T_{i}^{\mu y}\,\ln|x-y|\,\vec{I}_{i},
\end{equation}
In this equation, $G_{x,y}$ is the Green function of the
Laplacian:
$\nabla_{\bar{x}}^{-2}\delta^{(2)}(\bar{x}-y)=\frac{1}{2\pi}
\ln|x-y|\equiv G_{x,y}$.

Substituting \eref{2.11} in (\ref{2.7d}) we obtain the
area-invariant corresponding to the $0-th$ order
\begin{equation}\label{2.11d}
S^{(0)}_{YM}\Big|_{On-Shell}=-\frac{1}{\pi}\sum_{i,j}[\vec{I}_{i}\cdot\vec{I}_{j}]
\left(T_{i}^{\mu x}\,\ln|x-y|\,T_{j}^{\mu y}\right)\equiv
2\sum_{i,j}[\vec{I}_{i}\cdot\vec{I}_{j}]\;
J(\gamma_{i},\gamma_{j}),
\end{equation}
where we have defined the quantity $J(\gamma_{i},\gamma_{j})$. It
is worth noticing that this expression also corresponds to the
result that we would obtain if we had considered the Maxwell
theory coupled with Abelian particles. In section \ref{sec3} we
shall interpret the meaning of this invariant.

Introducing  (\ref{2.11}) and (\ref{2.9}) into (\ref{2.8})  we can
also obtain the first order Yang-Mills on-shell action in terms of
the curves $\gamma_{i}$. The result is
\begin{eqnarray}\label{2.12}
S^{(1)}_{YM}\Big|_{On-Shell}=\left(\frac{1}{2\pi}\right)^{2}\sum_{i,j,k}
[(\vec{I}_{i}\times\vec{I}_{j})\cdot\vec{I}_{k}]&&\left\{\frac12
T_{i}^{\mu x_{1}} T_{j}^{\nu x_{2}}T_{k}^{\nu
y}\ln|x-x_{1}|\ln|x-x_{2}|\partial_{\mu}\ln|x-y|
\right.\nonumber\\&&-\left.T_{i}^{\nu x\,\mu y}T_{j}^{\mu
x_{1}}T_{k}^{\nu x_{2}}\ln|y-x_{1}|\ln|x-x_{2}|\right\}.
\end{eqnarray}

\section{Interpretation of the invariants. Consistency
Conditions}\label{sec3}
\subsection{Interpretation of the $0-th$ order invariant $J(\gamma_{i}, \gamma_{j})$}

Since the isovectors $\vec{I}_{i}$ are arbitrary we conclude, from
(\ref{2.11d}), that the quantities
\begin{equation}\label{3.1}
J(\gamma_{i},\gamma_{j})\equiv -\frac{1}{2\pi}T_{i}^{\mu
x}\,\ln|x-y|\,T_{j}^{\mu
y}=-\frac{1}{2\pi}\int_{\mathbf{R}^{2}}d^{2}x\int_{\mathbf{R}^{2}}d^{2}y\,
T_{i}^{\mu \bar{x}}\,\ln|\bar{x}-\bar{y}|\,T_{j}^{\mu \bar{y}}
\end{equation}
must be area invariants. In order to interpret them we introduce
the $0$-form
\begin{equation}\label{3.2}
F_{y,\,\Sigma}\equiv
F(y,\Sigma)\equiv\int_{\Sigma}d^{2}x\,\delta^{(2)}(x-y),
\end{equation}
with support on the $2$-dimensional surface $\Sigma$ bounded by
$\gamma$. It can be readily seen that
\begin{equation}\label{3.3}
\varepsilon^{\mu\nu}\partial_{\nu
\bar{y}}F(\bar{y},\Sigma)=\int_{\gamma=\partial\Sigma}dx^{\mu}
\,\delta^{(2)}(x-y).
\end{equation}
Substituting  the  loop coordinates $T^{\mu x}$ in (\ref{3.1})
with the help of the previous expression we obtain
\begin{eqnarray}\label{3.5}
J(\gamma_{i},\gamma_{j})&=&-\frac{1}{2\pi}\int_{\mathbf{R}^{2}}d^{2}x\int_{\mathbf{R}^{2}}d^{2}y\,
\varepsilon_{\mu\alpha}\partial^{\alpha
\bar{y}}F_{\bar{y},\Sigma_{i}}\,\ln|\bar{x}-\bar{y}|\,\varepsilon^{\mu\beta}\partial_{\beta
\bar{x}}F_{\bar{x},\Sigma_{j}}\nonumber\\&=&-\frac{1}{2\pi}\int_{\mathbf{R}^{2}}d^{2}x\int_{\mathbf{R}^{2}}d^{2}y\,
\nabla^{2}_{\bar{x}}\,\ln|\bar{x}-\bar{y}|\,F_{\bar{y},\Sigma_{i}}\,F_{\bar{x},\Sigma_{j}}\nonumber\\&=&
\int_{\mathbf{R}^{2}}d^{2}x\,F_{\Sigma_{i}}(\bar{x})\,F_{\Sigma_{j}}(\bar{x})\nonumber\\
&=&\int_{\Sigma_{i}}d^{2}z\int_{\Sigma_{j}}d^{2}y\,\delta^{(2)}(z-y).
\end{eqnarray}

Expression (\ref{3.5}) measures the common area of the surfaces
$\Sigma_{i}$ and $\Sigma_{j}$ bounded by the closed curves
$\gamma_{i}$ and $\gamma_{j}$ respectively. Clearly, this quantity
is invariant under area-preserving diffeomorphisms, as expected.
It is interesting to notice that $J(\gamma_{i},\gamma_{j})$ is
formally analogous to the Gauss linking number of two curves in
$R^{3}$ \cite{Rolfsen}, which is just the $0-th$ order link
invariant obtained when the perturbative procedure discussed here
is applied to the classical Chern-Simons theory \cite{L3}.

Unfortunately, the ``trick'' of substituting the loop coordinate
$T^{\mu x}$ using \eref{3.3} and integrating by parts does not
suffice to yield a geometric interpretation of the first order
area-invariant (\ref{2.12}), since it is not possible to get rid
of the logarithm functions that appear in the expression for that
invariant. Later, in subsection \ref{c}, we shall present a
different method that provides a geometric interpretation of the
first order invariant, and hopefully of the higher order ones too.
Before, we study the consistence of the $0-th$ and first orders
equations of motion, since that method relies on the existence of
their solutions.

\subsection{Consistency conditions}\label{sec4}

The consistency of the $0-th$ order equation of motion
(\ref{2.10}) is immediate: taking the divergence of both sides of
the equation, and using our gauge condition, we obtain
\begin{equation}\label{2.13}
\partial_{\mu}\Box \vec{a}^{(0)\mu x}=\sum_{i}\partial_{\mu}T_{i}^{\mu x}\vec{I}_{i}=0,
\end{equation}
since  we are taking the trajectories of the Wong particles as
closed curves.

Regarding the first order equation  (\ref{2.16a}), we rewrite it
as
\begin{eqnarray}\label{2.14}
\Box \vec{a}^{(1)\nu x}=\vec{a}^{(0)}_{\mu
\bar{x}}\times\partial^{\nu}a^{(0)\mu \bar{x}}-2\vec{a}^{(0)\mu
\bar{x}}\times\partial_{\mu}\vec{a}^{(0)\nu
\bar{x}}+\vec{J}^{(1)\nu x}.
\end{eqnarray}
Taking the divergence on both sides of (\ref{2.14}) and using the
gauge condition we obtain
\begin{eqnarray}\label{2.15}
\partial_{\nu}\vec{a}^{(0)}_{\mu \bar{x}}\times\partial^{\nu}a^{(0)\mu
\bar{x}}-2\partial_{\nu}\vec{a}^{(0)\mu
\bar{x}}\times\partial_{\mu}\vec{a}^{(0)\nu \bar{x}}+\vec{a}_{\mu
\bar{x}}^{(0)}\times\Box \vec{a}^{(0)\mu
\bar{x}}+\partial_{\nu}\vec{J}^{(1)\nu x}=0.
\end{eqnarray}
Using expressions (\ref{2.9}), (\ref{2.10}) and (\ref{2.11}) for
$\vec{J}^{(1)\nu x}$, $\Box \vec{a}^{(0)\mu x}$ and
$\vec{a}^{(0)}_{\mu x}$ respectively, and taking into account the
differential constraint (\ref{2.6}) we have, after some
calculations
\begin{eqnarray}\label{2.16}
\sum_{i,j}\left(\vec{I}_{j}\times\vec{I}_{i}\right)\,\delta^{(2)}(x-x_{i}(0))\,
T_{i}^{\mu x}\,\ln|x-y|\,T_{j}^{\mu y}=0,
\end{eqnarray}
where $x_{i}(0)$ is the starting point of the curve $\gamma_{i}$.
If the curves do not intersect each other, and their iso-vectors
$\vec{I_{i}}$ are independent, equation (\ref{2.16}) leads to the
condition
\begin{equation}\label{2.20}
J(\gamma_{i},\gamma_{j})=0,
\end{equation}
for every $i,j$. Hence, the first order equation of motion is
consistent, and consequently the first order contribution to the
action is meaningful (for arbitrary values of the iso-vectors
$\vec{I_{i}}$) just when the $0-th$ order invariant
$J(\gamma_{i},\gamma_{j})$ vanishes for every $i,j$.

It is interesting to notice that this result resembles what occurs
in the study of link invariants in $R^{3}$ through this method
\cite{Nudos}: the next order invariant is meaningful whenever the
preceding one vanishes. In turn, this coincides with what happens
with the Milnor link invariants \cite{milnor}, which suggests that
the chain of area invariants that one would obtain following this
method could be seen as a kind of ``projection" of the Milnor
invariants, from links in $R^{3}$ and diffeomorphisms, to surfaces
in $R^{2}$ and area preserving diffeomorphisms.

\subsection{Axial gauge and bundle of parallel curves}\label{c}

Let us assign a straight line to every point $x$ of space. For
definiteness, take these lines starting at the spatial infinity,
running parallel to the vertical direction from bottom to top
until each one reaches its associated point $x$. The line
associated to $x$ is denoted by $\lambda^{x}$. The one-index
loop-coordinate $T^{\mu}(x,\lambda^{y})$ of $\lambda^{y}$ is
\begin{eqnarray}\label{form}
T^{\mu}(x,\lambda^{y})=\int_{\lambda^{y}}
dz^{\mu}\delta^{(2)}(x-z).
\end{eqnarray}

From the very structure of the perturbative equations of motion
(see comment after equation (\ref{2.16a})) $\Box a^{\mu (p)}=
f^{\mu}$, and the Lorentz gauge $\partial_{\mu}a^{\mu (p)}=0$, it
is clear that the Green function $G_{x,y}$ enters in the solution
of these equations only through its gradient
$\partial_{\bar{x}\mu} G_{\bar{x},y}$. Now, it is easy to see that
the substitution
\begin{eqnarray}\label{3.6a}
\partial^{\mu \bar{x}}G_{\bar{x},y}\quad\longrightarrow\quad
T^{\mu}(x,\lambda^{y}),
\end{eqnarray}
induces an Abelian gauge transformation on the potential $a^{\mu
x}$
\begin{eqnarray}\label{3.7a}
a^{\mu (p)}\quad\longrightarrow\quad a_{\psi}^{\mu (p)} = a^{\mu
(p)}+\partial^{\mu }\psi,
\end{eqnarray}
where $\psi$ is a certain function depending on the bundle of
curves $\lambda$. Moreover, it can be shown that if $\eta^{\mu}$
is a vector in the direction of $\lambda$, $a_{\psi}^{\mu
(p)}\eta_{\mu}=0$. Hence, the replacement \eref{3.6a} amounts to
changing from Lorentz  to axial gauge [actually, since space is
Euclidean, Lorentz and Coulomb gauges are identical, as well as
axial and temporal gauges]. Finally, it can be shown that (at
least) the $0-th$ and first order contributions to the action on
shell are invariant under Abelian gauge transformations, provided
that their respective equations of motion are consistent. Thus we
arrive to the following result. Although the calculations are done
in the Lorentz gauge, we are allowed to perform an Abelian gauge
transformation whenever the consistence conditions holds (it seems
that this result holds to every order, but we do not have a proof
yet). Changing to the axial gauge will allow us to interpret the
$0-th$ and first order invariants. Let us begin with the former.

Starting from expression (\ref{3.1}) one has
\begin{eqnarray}\label{3.8a}
J(\gamma_{i},\gamma_{j})&=&-\frac{1}{2\pi}\epsilon^{\mu\alpha}\partial_{\alpha
x}F_{\Sigma_{i}x}T_{j}^{\mu
y}\ln|x-y|\nonumber\\&=&\epsilon^{\mu\alpha}F_{\Sigma_{i}x}T_{j}^{\mu
y}\partial_{\alpha x}G_{x,y}\nonumber\\&=&
\epsilon^{\mu\alpha}F_{\Sigma_{i}x}T_{j}^{\mu
y}T_{\alpha}(x,\lambda^{y}).
\end{eqnarray}

\begin{figure}[!hbt]
\centerline{\includegraphics[angle=14,height=8.0cm,
width=3.5in]{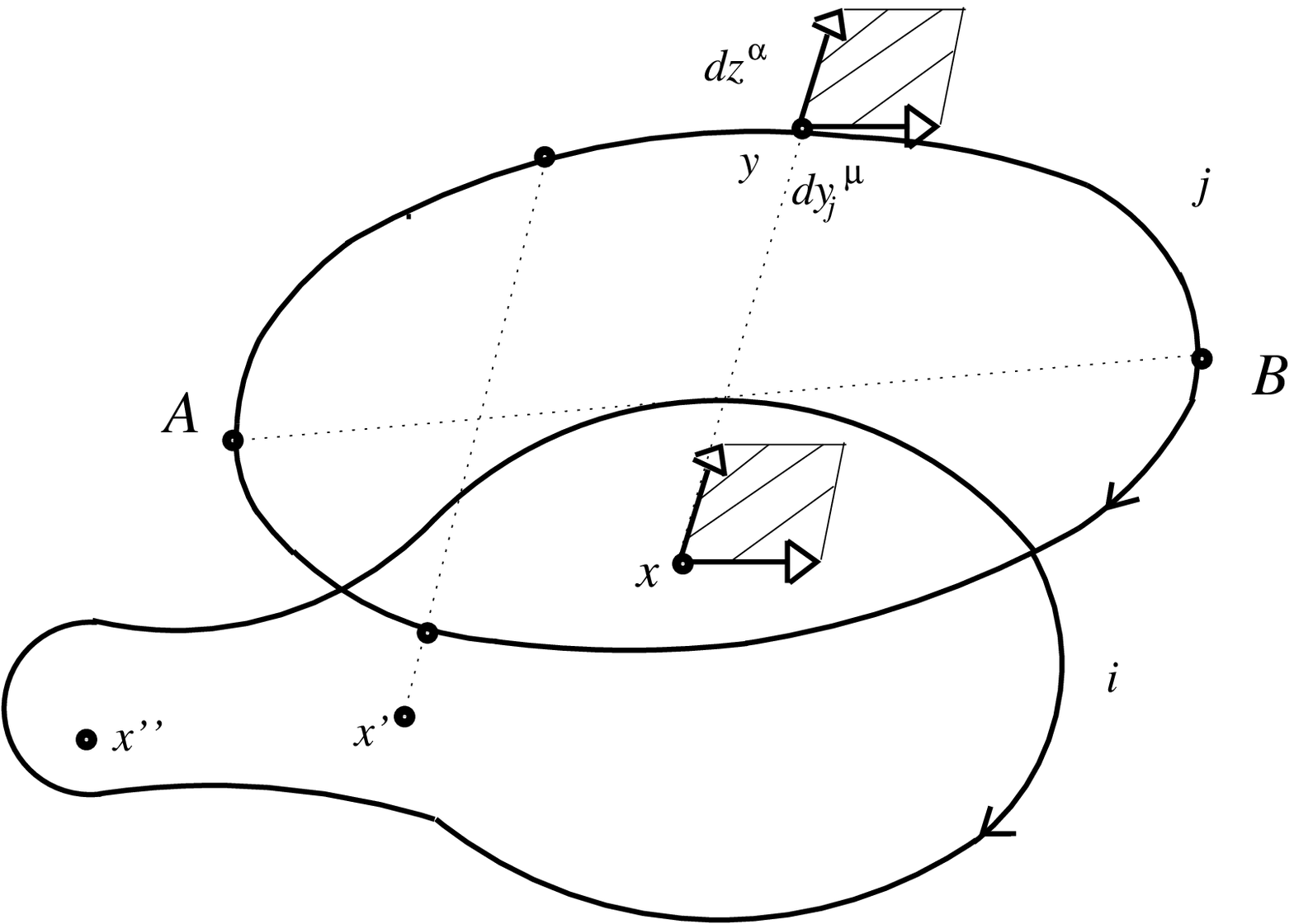}}
\caption{Configuration with a non-vanishing $0-th$ order
contribution}\label{F1}
\end{figure}

Now, observe that the integration in $x$ must be restricted to the
area surrounded by curve $i$, due to the factor $F_{\Sigma_{i}x}$
in the integrand. Furthermore, for each $x$, the presence of the
factor $T_{\alpha}(x,\lambda^{y})$ ensures that there will be
contributions  only when this point $x$ is connected through a
vertical line with a point $y$ lying in the other curve $j$, in
such a manner that the infinitesimal displacements along the
vertical path ($dz^{\alpha}$) and along the closed curve $j$
($dy^{\mu}$) form a non-degenerated parallelogram (see FIG.
\ref{F1}). Under these conditions, the contribution associated to
the point $x$ is just the area of that infinitesimal
parallelogram: $\epsilon^{\mu\alpha} dy^{\mu} dz^{\alpha}$.

On the other hand, for each $x$ belonging to $\Sigma_{i}$, but not
to the intersection (see $x^{,}$ in FIG. \ref{F1}), there would be
two contributions  (more generally, an even number of them, which
could be none as in the case of $x^{,,}$, shown in FIG. \ref{F1})
because the parallel curves intersect twice the closed trajectory
$j$. But in these cases, the net contribution of the ``upper side"
of curve $j$ (i.e. the piece from $A$ to $B$ in clockwise sense)
plus the ``down side'' (the part from $B$ to $A$ in
counterclockwise sense) cancel each other.

Hence, we have to sum up as many infinitesimal parallelograms as
necessary to cover the area of intersection, and we recover our
previous interpretation of the $0-th$ order invariant.

\subsection{Interpretation of the $1-st$ order invariant $S^{(1)}$ and final remarks.}

Our starting point will be  expression (\ref{2.12}) for the
Yang-Mills action on-shell up to first order, expressed entirely
in terms of the closed curves. Since the coefficient related to
the  iso-vectors $\vec{I}_{i}$ is completely antisymmetric in
$i,j,k$, and these iso-vectors are arbitrary, we conclude that the
following quantity is invariant:
\begin{eqnarray}\label{3.8}
Y(\gamma_{i},\gamma_{j},\gamma_{k})\equiv\ &&
\left(\frac{1}{2\pi}\right)^{2}\left(\frac12 \, T_{[i}^{\mu x_{1}}
T_{j}^{\nu x_{2}}T_{k]}^{\nu
y}\ln|x-x_{1}|\ln|x-x_{2}|\partial_{\mu}\ln|x-y|
 \nonumber\right.\\&&\left. - T_{[i}^{\nu x\,\mu y}T_{j}^{\mu
x_{1}}T_{k]}^{\nu x_{2}}\ln|y-x_{1}|\ln|x-x_{2}|\right).
\end{eqnarray}
Here, square brackets denote anti-symmetrization.

Consider the first term in (\ref{3.8}). Making the substitution
(\ref{3.3}), and after some calculations it results
\begin{eqnarray}\label{3.9}
&&T_{i}^{\mu x_{1}} T_{j}^{\nu x_{2}}T_{k}^{\nu
y}\ln|x-x_{1}|\ln|x-x_{2}|\partial_{\mu
x}\ln|x-y|\nonumber\\
&&=\epsilon^{\mu\alpha}\partial_{\alpha}F_{\Sigma_{i}x_{1}}\epsilon^{\nu\beta}\partial_{\beta}F_{\Sigma_{j}x_{2}}
\epsilon^{\nu\gamma}\partial_{\gamma}F_{\Sigma_{k}y}\ln|x-x_{1}|\ln|x-x_{2}|\partial_{\mu
x}\ln|x-y|\nonumber\\&&=-\epsilon^{\mu\alpha}F_{\Sigma_{i}x_{1}}F_{\Sigma_{j}x_{2}}F_{\Sigma_{k}y}
\partial_{\alpha x_{1}}\ln|x-x_{1}|\partial_{\beta x_{2}}\ln|x-x_{2}|
\partial_{\beta y}\partial_{\mu x}\ln|x-y|\nonumber\\
&&=-(2\pi)^{3}\epsilon^{\mu\alpha}F_{\Sigma_{i}x_{1}}F_{\Sigma_{j}x_{2}}F_{\Sigma_{k}y}
\partial_{\alpha x_{1}}G_{x,x_{1}}\partial_{\beta x_{2}}G_{x,x_{2}}
\partial_{\beta y}\partial_{\mu x}G_{x,y}.
\end{eqnarray}
Making the substitution (\ref{3.6a}), i.e., changing to axial
gauge, the last line of (\ref{3.9}) can  be written down as
\begin{eqnarray}\label{3.10}
-(2\pi)^{3}\epsilon^{\mu\alpha}F_{\Sigma_{i}x_{1}}F_{\Sigma_{j}x_{2}}F_{\Sigma_{k}y}
T_{\alpha}(x, \lambda^{x_{1}})T_{\beta}(x, \lambda^{x_{2}})
\partial_{\beta y}T_{\mu}(x, \lambda^{y}).
\end{eqnarray}
This quantity vanishes, because the straight lines $\lambda^{x}$
are parallel, which implies that the cross product of their
tangent vectors is zero.

It remains to compute the second term in (\ref{3.8}). Changing to
the axial gauge by means of substitution (\ref{3.6a}), we obtain
\begin{eqnarray}\label{3.11}
Y(\gamma_{i},\gamma_{j},\gamma_{k})= - T_{i}^{\nu x,\,\mu
y}\epsilon^{\mu\alpha}\epsilon^{\nu\beta}F_{\Sigma_{j}x_{1}}F_{\Sigma_{k}x_{2}}T_{\alpha}(y,
\lambda^{x_{1}})T_{\beta}(x, \lambda^{x_{2}}) + [ijk],
\end{eqnarray}
where $[ijk]$ means that the whole expression must be
anti-symmetrized in the indices inside the bracket. This symmetry
property of \eref{3.11} imply that the minimal number of curves
that will produce a non-trivial result is three.

\begin{figure}[!hbt]
\centerline{\includegraphics[width=3.5in]{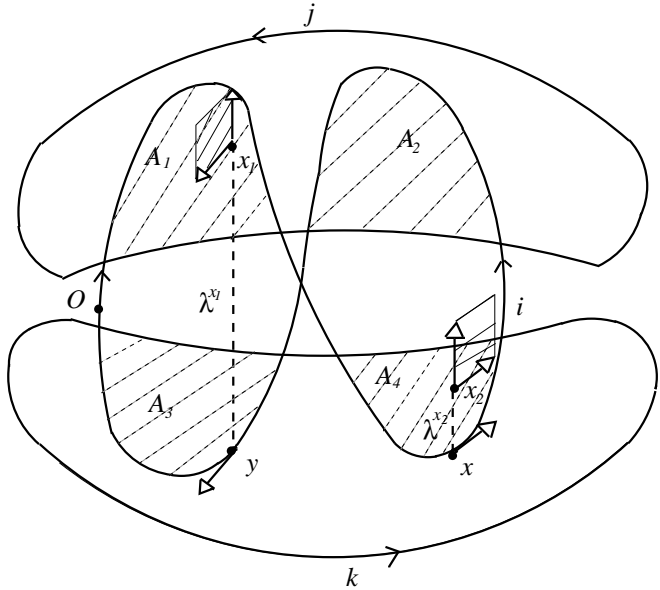}}
\caption{Example of a ``linking-area'' configuration detected by
$Y(\gamma_{i},\gamma_{j},\gamma_{k})$}\label{F2}
\end{figure}

To interpret this expression, let us then consider a configuration
of three curves as in FIG. \ref{F2}. This configuration was chosen
taking into account that, as we already know,
$Y(\gamma_{i},\gamma_{j},\gamma_{k})$ is meaningful whenever
(\ref{3.1}) vanishes for every pair of curves. In this
configuration, the absolute values of the areas of intersection
$A_{i}$, $i=1,2,3,4$ are the same, but their signs are such that
the above condition is fulfilled. It should be noticed that, given
the configuration of FIG. \ref{F2}, only the first term in
\eref{3.11} contributes.
The points $x_{1}$ and $x_{2}$ must belong to the surfaces
$\Sigma_{j}$ and $\Sigma_{k}$ respectively, as indicated by the
surface-functions $F_{\Sigma_{j}x_{1}}$ and $F_{\Sigma_{k}x_{2}}$.
Also, we  have now two bundles of open straight parallel lines:
the ``fibers'' of one of them (that corresponding to
$T_{\alpha}(y, \lambda^{x_{1}})$) start at the points $y$  of
$\gamma_{i}$, and end at the points $x_{1}$ of the surface
surrounded by $\gamma_{j}$. The other bundle of parallel lines
(corresponding to $T_{\beta}(x, \lambda^{x_{2}})$) also begins at
$\gamma_{i}$, but finishes at $x_{2}$ inside $\Sigma_{k}$. As
before, the contributions to expression \eref{3.11} are expressed
as areas of the infinitesimal parallelograms formed by vertical
displacements along both the fiber and the curve where the
corresponding fiber ends.

Up to this point, there is nothing specially new in the discussion
of the first-order invariant. The main difference with the
previous case, is due to the presence of the loop coordinate with
two indices $T_{i}^{\nu x,\,\mu y}$, which introduces an
\emph{intrinsic} order in curve $i$ that affects which
contributions must be summed up. In the integration along curve
$i$ the areas of the parallelograms coming from curve $k$, which
have to be multiplied by those coming from curve $j$, contribute
only if the former are ``created'' before the later when one
travels along curve $i$ starting at an arbitrary marked point
($O$, in FIG.\ref{F2}). Henceforth, of the four products of areas
$|A_{1}||A_{4}|$, $-|A_{1}||A_{3}|$, $|A_{2}||A_{3}|$ and
$-|A_{2}||A_{4}|$, that would appear if instead of $T_{i}^{\nu
x,\,\mu y}$ we had $T_{i}^{\nu x}$ times $T_{i}^{\mu y}$ in
\eref{3.11}, the last one does not contribute due to the fact that
$A_{2}$ is ``drawn" after $A_{4}$ was.

Then, the result of applying \eref{3.11} to the picture in FIG.
\ref{F2} is minus the square of the area $A_{i}^{2}$ of any of the
lobules of intersection. Clearly, this is an area-preserving
diffeomorphism invariant, which could not be detected by the
previous order one (which in fact vanishes).

Summarizing, we have shown that classical $2d$ Yang-Mills theory
coupled to non-dynamical Wong particles in Euclidean space can be
used to obtain area-invariants through a perturbative scheme. This
fact may be seen as an instance of a general result established in
\cite{rafael}, which in turn aroused as a formalization of
previous results \cite{L3,Nudos}.

The first two invariants provided by this method were explicitly
computed, and a geometrically appealing interpretation was given.
It is conjectured that the sequence of invariants here obtained
could be seen as a kind of ``projection" of the link-invariants of
Milnor \cite{milnor}. This and other pertinent questions will be
addressed in future work.\\

We thank Edmundo Castillo for his assistance with the figures.
This work was partially supported  by  Fonacit grant G2001000712.

\end{document}